\documentclass[12pt,preprint]{aastex6}

\shorttitle{light wall, flare and light bridge}
\shortauthors{Hou et al.}

\begin{document}

\title{\textbf{A solar flare disturbing a light wall above a sunspot light bridge}}

\author{Yijun Hou\altaffilmark{1,2}, Jun Zhang\altaffilmark{1,2}, Ting Li\altaffilmark{1,2},
        Shuhong Yang\altaffilmark{1,2}, Leping Li\altaffilmark{1} and Xiaohong Li\altaffilmark{1,2}}

\altaffiltext{1}{Key Laboratory of Solar Activity, National Astronomical Observatories,
 Chinese Academy of Sciences, Beijing 100012, China; yijunhou@nao.cas.cn}

\altaffiltext{2}{University of Chinese Academy of Sciences, Beijing 100049, China}

\begin{abstract}
With the high-resolution data from the \emph{Interface Region Imaging Spectrograph},
we detect a light wall above a sunspot light bridge in the NOAA active region (AR)
12403. In the 1330 {\AA} slit-jaw images, the light wall is brighter than the ambient
areas while the wall top and base are much brighter than the wall body, and it keeps
oscillating above the light bridge. A C8.0 flare caused by a filament activation
occurred in this AR with the peak at 02:52 UT on 2015 August 28, and the flare's
one ribbon overlapped the light bridge which was the observational base of the
light wall. Consequently, the oscillation of the light wall was evidently disturbed.
The mean projective oscillation amplitude of the light wall increased from 0.5 Mm to
1.6 Mm before the flare, and decreased to 0.6 Mm after the flare. We suggest that
the light wall shares a group of magnetic field lines with the flare loops, which
undergo a magnetic reconnection process, and they constitute a coupled system.
When the magnetic field lines are pushed upwards at the pre-flare stage, the light
wall turns to the vertical direction, resulting in the increase of the light wall's
projective oscillation amplitude. After the magnetic reconnection takes place,
a group of new field lines with smaller scales are formed underneath the reconnection
site and the light wall inclines. Thus, the projective amplitude decreases remarkably
at the post-flare stage.
\end{abstract}


\section{Introduction}
Solar flares are energetic phenomena in the solar atmosphere, releasing dramatic
electromagnetic energy spanning the range from X-ray to radio wavelengths. In the
standard two-dimensional (2D) flare model (CSHKP models; Carmichael 1964; Sturrock 1966;
Hirayama 1974; Kopp \& Pneuman 1976), a filament rises above the neutral line and
then initially drives the flare process. The rising filament pushes the overlying
magnetc field lines upwards, and the resulting losses of pressure below form an inward
magnetic force towards the neutral sheet. This force drives antiparallel magnetic
filed lines to converge, leading to the formation of a current sheet, and magnetic
reconnection begins to take place. Thus the released energy heats the coronal plasma
and also accelerates particles. The accelerated particles flow downwards from the
reconnection site along the newly formed magnetic field lines, and in the lower solar
atmosphere, the flare ribbons are generated (Priest \& Forbes 2002). The flare ribbons
observed in H$\alpha$ and ultraviolet (UV) wavelengths are the conspicuous characteristics
of solar flares and are usually located on either sides of the polarity inversion line.
The flare ribbons move apart during the reconnection process and the separation generally
stops at the edge of the sunspots. However, Li \& Zhang (2009) reported that flare
ribbons sometimes sweep across the whole sunspots.

Sunspots are concentrations of magnetic fields and the overturning motion of the plasma
is hindered by the strong magnetic field in the sunspot umbra (Gough \& Tayler 1966).
Bright structures within the umbra are signatures of not completely suppressed
convection, and light bridges are the best known representatives of these structures
(Sobotka et al. 1993; Borrero \& Ichimoto 2011). The magnetic field of light bridges
is generally weaker and more inclined than the local strong and vertical field
(Lites et al. 1991; Rueedi et al. 1995; Leka 1997; Jur{\v c}{\'a}k et al. 2006).
Recent simulations and observations have shown that a light bridge's magnetic field
is twisted and related to emerging magnetic fields (Louis et al. 2015; Toriumi et al.
2015a,b; Yuan \& Walsh 2016). Above the light bridges, some chromospheric activities
have been observed in the forms of jets and surges (Asai et al. 2001; Shimizu et al. 2009;
Louis et al. 2014). Recently, Yang et al. (2015) reported an oscillating light wall above a
sunspot light bridge. The light wall is brighter than the ambient areas while the wall top
is much brighter than the wall body in 1330 {\AA}. Hou et al. (2016) revealed that some
light walls are multilayer and multithermal structures which occur along magnetic
neutral lines in active regions (ARs), not just above the light bridge. However, the work
about the magnetic topology of the light wall is rare.

In this Letter, we report that a light wall (oscillating above a sunspot light bridge)
is disturbed by a C8.0 flare while one ribbon of this flare intrudes into the sunspot and
overlaps the light bridge. The mean projective amplitude of light wall's oscillation
increases at the pre-flare stage, and decreases after the flare. Using the coordinated
observations from the \emph{Interface Region Imaging Spectrograph} (\emph{IRIS};
De Pontieu et al. 2014) and the \emph{Solar Dynamic Observatory} (\emph{SDO};
Pesnell et al. 2012), we investigate this event in detail for understanding the
magnetic configuration of the light wall.

\section{Observations and Data Analysis}
On 2015 August 28, a C8.0 flare took place in NOAA AR 12403 with one ribbon intruding into
the main sunspot of the AR. Moreover, the flare ribbon overlapped a light bridge inside
the sunspot and disturbed the oscillation of the light wall above this light bridge.
From 23:09:50 UT on 2015 August 27 to 03:48:25 UT on August 28 , the \emph{IRIS} was
pointed at AR 12403, and clearly observed the light wall, flare and light bridge.
We hence obtain a series of \emph{IRIS} slit-jaw 1330 {\AA} images (SJIs)
with a cadence of 18 s, a pixel scale of 0.{\arcsec}333, and a field of view (FOV)
of 120{\arcsec} $\times$ 119{\arcsec}. The 1330 {\AA} channel contains emission
from the strong C II 1334/1335 {\AA} lines that are formed in the upper chromosphere
and transition region. To check the coronal mass ejection (CME) associated with
this C8.0 flare, we also employ the images of the solar corona, taken by the Large
Angle and Spectrometric Coronagraph (LASCO) on board the \emph{Solar and Heliospheric
Observatory} (\emph{SOHO}; Brueckner et al. 1995).

Moreover, the observations of the Atmospheric Imaging Assembly (AIA; Lemen et al. 2012)
and the Helioseismic and Magnetic Imager (HMI; Scherrer et al. 2012) onboard the
\emph{SDO} are used as well. The AIA takes full-disk images in ten (E)UV channels with
a cadence of 12 s and spatial sampling of 0.{\arcsec}6 pixel$^{-1}$. We adopt the
observations of AIA 131 {\AA}, 94 {\AA}, 335 {\AA}, 211 {\AA}, 193 {\AA}, 171 {\AA},
1600 {\AA}, and 304 {\AA} on 2015 August 28 to investigate the coupled system consisting
of the light wall, flare, and light bridge. The full-disk line-of-sight (LOS) magnetograms
and the intensitygrams from the HMI with a cadence of 45 s and a sampling of 0.{\arcsec}5
pixel$^{-1}$ are also applied.

\section{Results}
The event of interest took place around the main sunspot of NOAA AR 12403 (see Figure 1 and
movie1.mp4). On 2015 August 28, the AR approached the solar south-western limb and its
overview is shown in Figures 1(a) and 1(b). Figures 1(c)-1(f) expand part of this region
(see the green square in panel (a)) and display the light wall, flare ribbons, and light
bridge in different wavelengths, as well as a photospheric magnetogram. Combining the
1330 {\AA} SJIs and HMI intensitygrams, we detect a
light wall (see the white arrow in panel (c)) above the sunspot light bridge (see the red
arrow in panel (e)). Checking the \emph{IRIS} and the \emph{SDO} observations, we notice
that this light wall oscillated above the light bridge from 2016 August 26 to August 29
until this region rotated to the farside of the Sun. On August 28, a C8.0 two-ribbon flare
took place near the main sunspot around 02:42 UT. We delineate the two ribbons
of the flare by blue and red dashed curves in 1600 {\AA} image of panel (d) and duplicate
them in panels (c), (e), and (f). The western flare ribbon (see the red dashed lines
in panels (c)-(e)) overlapped the light bridge which was the observational base of the light
wall. The HMI LOS magnetogram in panel (f) shows that the overlapping flare ribbon is located
in strong positive magnetic fields of the sunspot while the eastern ribbon in plage region
with negative magnetic fields (see the blue dashed lines in panels (d) and (f)).

To investigate the evolution of this C8.0 flare, we check the \emph{SDO}/AIA observations
from 02:28 UT to 03:48 UT on 2015 August 28. In AIA 304 {\AA} channel, a quiescent filament
lay above the magnetic neutral line before the flare's onset. Around 02:41 UT, this filament
was partly activated (see Figure 2(a)). The C8.0 flare started at 02:42 UT and immediately
at the onset of the flare, the flare loops exhibited apparent slipping motions. 
\textbf{This phenomenon is similar to the observations of Dud{\'{\i}}k et al. (2014, 2016) 
and Li \& Zhang (2015), implying the occurrence of slipping magnetic reconnection.} 
The brightening of the eastern footpoints of the flare loops
gradually propagated towards the southeast, which developed into the eastern flare ribbon. 
Meanwhile, the brightening of the flare loops' western footpoints moved to the northwest 
(see the green arrows in panel (b)), forming the western flare ribbon. At 02:52 UT, the 
flare increased to its maximum and the two flare ribbons appeared on either side of the 
neutral line (see panel (b)). During the next ten minutes, a dark filament material flow 
from the northwest to the southeast was detected between the two flare ribbons (\textbf{see the black
line and arrow in panel (b)}). Thus, we consider the activation (or eruption) of the filament as the cause
of this flare. The hot flare loops were observed well in AIA 94 {\AA} passband. We delineate
these loops with black solid, dashed and dotted lines in panel (d) and duplicate them to
panel (c). Using the differential emission measure analysis method which is based on the
``xrt\_dem\_iterative2.pro" in the Solar Software package (Cheng et at. 2012), we obtain the
temperature map as displayed in panel (e). The temperature of the flaring region increased
significantly and the outlines of hot loops are also obvious in panel (e). At the top of the
flare loops (see the black square in panel (e)), the temperature reaches the maximum of
about 9.0 MK. Moreover, the LASCO C2 observations show that a CME is associated with the
filament activation and this flare. The CME ejected with an average speed of 253 km s$^{-1}$
and a width angle of 21{\degr} (see the \textbf{LASCO} C2 difference image in panel (f)). The white
plus symbol in panel (f) roughly marks the flare's location in the solar surface.

When the flare occurred, its one ribbon intruded into the sunspot and overlapped the light
bridge which was the observational base of the oscillating light wall. To study the flare's
disturbance to this light wall, temporal evolution of the light wall is examined.
We cut out a smaller FOV in \emph{IRIS} 1330 {\AA} SJIs (see the white square in Figure
1(c)) and rotate it 130{\degr} anticlockwise, which are shown in Figures 3(a)-3(f).
The panels (a)-(c) exhibit the light wall's oscillation before the onset of flare.
In the 1330 {\AA} channel, the light wall is brighter than the ambient areas while its
top and base are brighter than the wall body. One of the flare's ribbons brightened
successively, and approached the light wall around 02:46 UT (see panel (c)). Then this
flare ribbon overlapped the light bridge and light wall completely between 02:47 UT and
02:51 UT. After that, the light wall kept oscillating above the light bridge (see panels
(d)-(f)). But the light wall's oscillation was disturbed (see movie2.mp4). For more
details about the disturbance, we make a space-time plot along the slice ``A-B" marked
in panels (a)-(f) and display it in panel (g). The
blue solid curve outlines the position evolution of the light wall top, and the blue
dashed line marks the light wall base. The mean period of the oscillation is about 4.0
minutes. It is shown that after the T1 (02:13 UT) in panel (g), that is 30 minutes before
the flare ribbon's appearance, the oscillation amplitude of the light wall increased
evidently and maintained at a high level until the flare ribbon approached the light
wall at T2 (02:46 UT). Then the light wall oscillated with a smaller
amplitude. These two time points (T1 and T2) divide the whole process into three phases
and we estimate an average projective maximum height and oscillation amplitude of the
light wall for each phase. The distance between the wall base and the wall top is
calculated as the height of the light wall. And we calculate the mean projective maximum
height according to the formula: $\overline{H_{max}} = \frac{\sum_{0}^{N-1} H_{max}^i}{N}$,
where $H_{max}^i$ is the projective height of wall top oscillation peak \emph{i} in one
cycle. Moreover, the mean projective amplitude of the light wall's oscillation is
$\overline{AMP}= \frac{\sum_{0}^{N-1} H_{max}^i-H_{min}^i}{2N}$, where $H_{max}^i$ and
$H_{min}^i$ are the projective heights of wall top oscillation peak \emph{i} and valley
\emph{i} in one cycle. The calculated average projective maximum heights of the light wall
in the three phases are 2.7 Mm, 5.2 Mm, and 3.0 Mm, respectively. The mean projective amplitudes
of the light wall's oscillation in three phases are separately 0.5 Mm, 1.6 Mm, and 0.6 Mm.

\section{Summary and Discussion}
With the high tempo-spatial \emph{IRIS} and \emph{SDO} observations, we detect a light
wall oscillating with a mean period of 4.0 minutes above a sunspot light bridge in NOAA
AR 12403. The light wall is brighter than the ambient regions while the top and base
of the light wall are much brighter than the wall body in 1330 {\AA} channel. On 2015
August 28, a C8.0 flare caused by a filament's activation occurred in this AR with the
peak at 02:52 UT. We first observe that one of
the flare ribbons intruded into the sunspot and then overlapped the light bridge which was
the observational base of the light wall. As a result, the oscillation of light wall
was obviously disturbed by the ribbon. The mean projective oscillation amplitude of
the light wall increased from 0.5 Mm to 1.6 Mm before the flare, and decreased to 0.6
Mm after the flare. In addition, the images of the LASCO C2 on board the \emph{SOHO}
are adopted to study the CME related to the C8.0 flare. This flare is an eruptive
flare which results in a CME with an average speed of 253 km s$^{-1}$ and a width
angle of 21{\degr}.

The light wall has been reported in several works. Yang et al. (2015) reported an
oscillating light wall above a sunspot light bridge and interpreted the oscillations
of the light wall as the leakage of p-modes from below the photosphere. Hou et al.
(2016) revealed that some light walls are multilayer and multithermal structures
which occur along magnetic neutral lines in active regions. As a newfound structure,
the light wall's driving mechanism and magnetic topology have not been well understood.
The present work reports a C8.0 flare disturbing a light wall above a sunspot light bridge,
which may contribute to the understanding of the light wall's magnetic configuration.
Louis et al. (2014) proposed that the dynamic chromospheric jets above the light bridge
seem to be guided by the magnetic field lines. Tian et al.(2014) reported sub-arcsecond
bright dots in the transition region above sunspots and suggested that some of these
bright dots appear to be located at the bases of magnetic loops. In this event, we
suggest that the light wall shares a group of magnetic field lines with the flare loops
involved in magnetic reconnection process, and is located at the bases of these magnetic
loops. They constitute a coupled system (see Figure 4(a)). When the activated filament
begins to rise, the overlying field lines, which connect the positive light bridge fields
and the negative plage fields, are pushed upwards before the flare's onset (between T1
and T2 in Figure 3(g)). Then the light wall turns to close to the vertical direction
(see Figure 4(b)). Due to the projection effect, the observations along the LOS show
the light wall's mean projective maximum height increases from 2.7 Mm to 5.2 Mm. Around
the T2 in Figure 3(g), magnetic reconnection takes place (see Figure 4(c)). Underneath the
reconnection site, the magnetic filed lines with smaller scales are newly formed and the
flare ribbons appear. As a result, the light wall turns to away from the vertical
direction (see Figure 4(d)). Thus, the average projective maximum height of light wall
decreases remarkably to 3.0 Mm after the flare (after the T2 in Figure 3(g)).

Checking the AIA 304 {\AA} observations during the flare's evolution, we detect a
quiescent filament lying above the neutral line. When the flare occurred, the
dark material flow was observed between the two flare ribbons. Therefore, we
consider that this filament was partly activated and subsequently rose, leading to
the C8.0 flare. For eruptive flares, the vertical magnetic fields on both sides of
the current sheet correspond to the legs of CME-related expanding field lines
(Forbes et al. 2006; Aulanier et al. 2010). Here, we detect a CME associated with this
flare by LASCO C2 data, which is consistent with the illustration in Figure 4.

\acknowledgments {\textbf{We thank the referee for his/her valuable suggestions.}
The data are used courtesy of \emph{IRIS}, \emph{SDO} and \emph{SOHO}
science teams. \emph{IRIS} is a NASA small explorer mission developed and operated by LMSAL
with mission operations executed at NASA Ames Research center and major contributions
to downlink communications funded by ESA and the Norwegian Space Centre. This work
is supported the National Natural Science Foundations of China (11533008, 11303050,
11673035 and 11673034) and the Strategic Priority Research Program$-$The Emergence
of Cosmological Structures of the Chinese Academy of Sciences (Grant No. XDB09000000).}

{}
\clearpage

\begin{figure}
\centering
\includegraphics [width=0.89\textwidth]{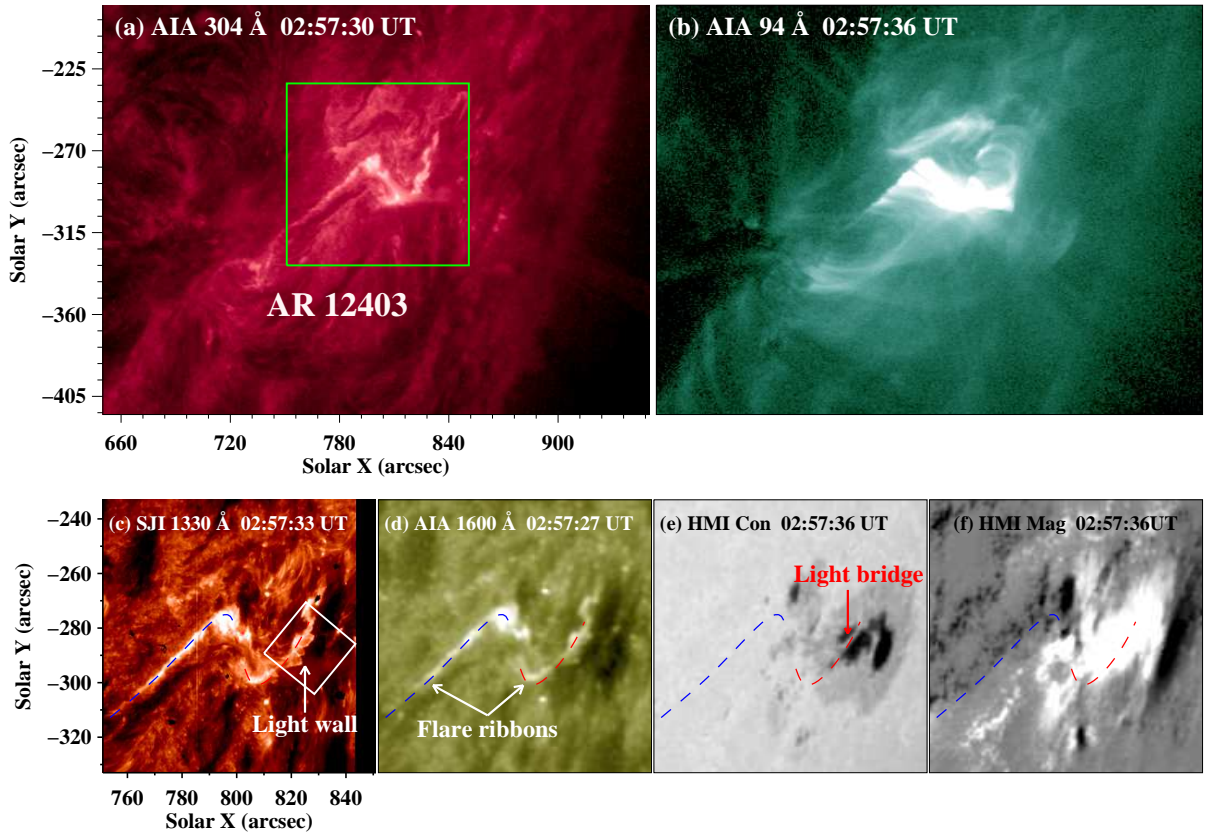}
\caption{
Panels (a)-(b): \emph{SDO}/AIA 304 {\AA} and 94 {\AA} images displaying the overview
of AR 12403 on 2015 August 28. The green square outlines the FOV of panels (c)-(f).
Panels (c)-(f): \emph{IRIS} 1330 {\AA} SJI, \emph{SDO}/AIA 1600 {\AA} image, HMI
continuum intensity and HMI LOS magnetogram showing the relative positions of the
light wall, light bridge and flare ribbon, as well as their underlying magnetic fields.
The blue and red dashed curves delineate two ribbons of the flare.
(An animation (movie1.mp4) of this figure is available.)
}
\label{fig1}
\end{figure}

\begin{figure}
\centering
\includegraphics [width=0.89\textwidth]{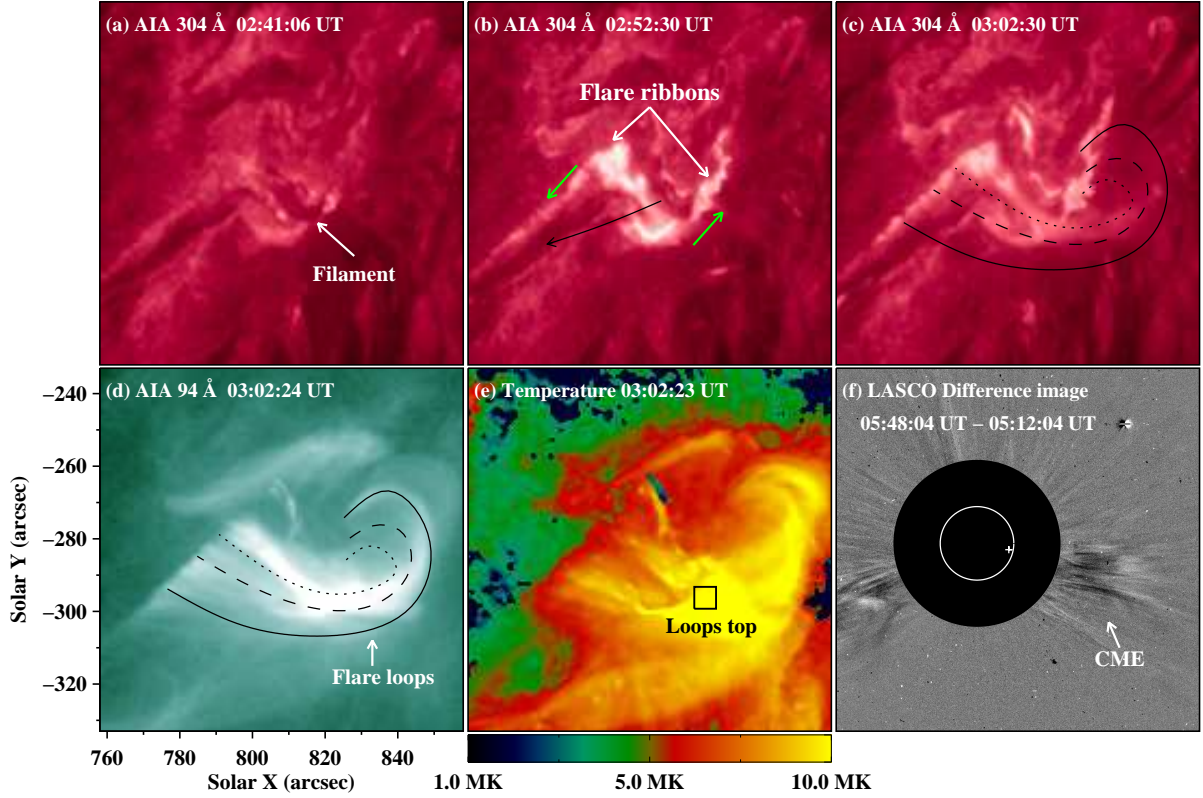}
\caption{
Panels (a)-(c): sequence of AIA 304 {\AA} images showing the flare caused by a filament's
activation. The green arrows in panel (b) mark the propagation directions of the brightenings
of the flare loops' footpoints. \textbf{The black line and arrow in panel (b) approximate 
the trajectory of the dark filament material flow.}
Panel (d): AIA 94 {\AA} image exhibiting the high-temperature flare loops. The black solid,
dashed, and dotted lines delineate these loops and they are duplicated to panel (c).
Panel (e): temperature map displaying the region's temperature after the flare. The black
square outlines the region around loops top. The FOV of panels (a)-(e) is outlined by
the green square in Figure 1(a).
Panel (f): LASCO C2 difference image showing the CME associated with the eruptive flare.
The white circle outlines the solar disk. And the white plus marks the flare's location
in the solar surface.
}
\label{fig2}
\end{figure}

\begin{figure}
\centering
\includegraphics [width=0.89\textwidth]{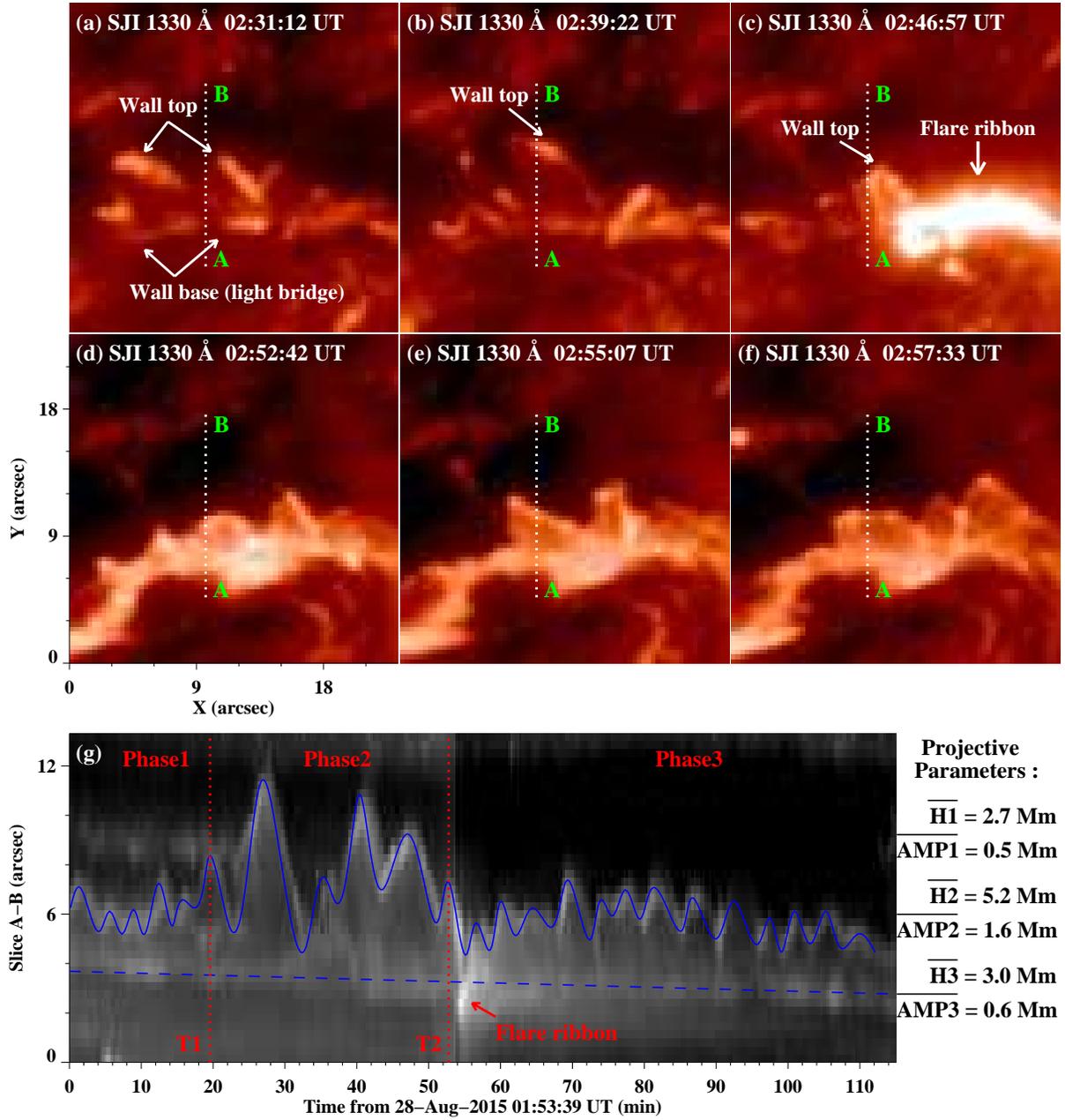}
\caption{
Panels (a)-(f): sequence of the \emph{IRIS} 1330 {\AA} SJIs showing the light wall before
and after the disturbance of the flare ribbon. Their FOV is outlined by the solid white
square in Figure 1(c), which has been rotated by -130{\degr}.
Panel (g): space-time plot along the slice ``A-B" marked in panels (a)-(f). The blue solid
curve delineates the light wall top, and the blue dashed line labels the wall base (light
bridge). Two time points (red vertical dotted lines) divide the whole process into three
phases. We give average projective maximum heights ($\overline{H1}$, $\overline{H2}$,
and $\overline{H3}$) and oscillation amplitudes ($\overline{AMP1}$, $\overline{AMP2}$, and
$\overline{AMP3}$) of the light wall for all the three phases.
(An animation (movie2.mp4) of this figure is available.)
}
\label{fig3}
\end{figure}

\begin{figure}
\centering
\includegraphics [width=0.89\textwidth]{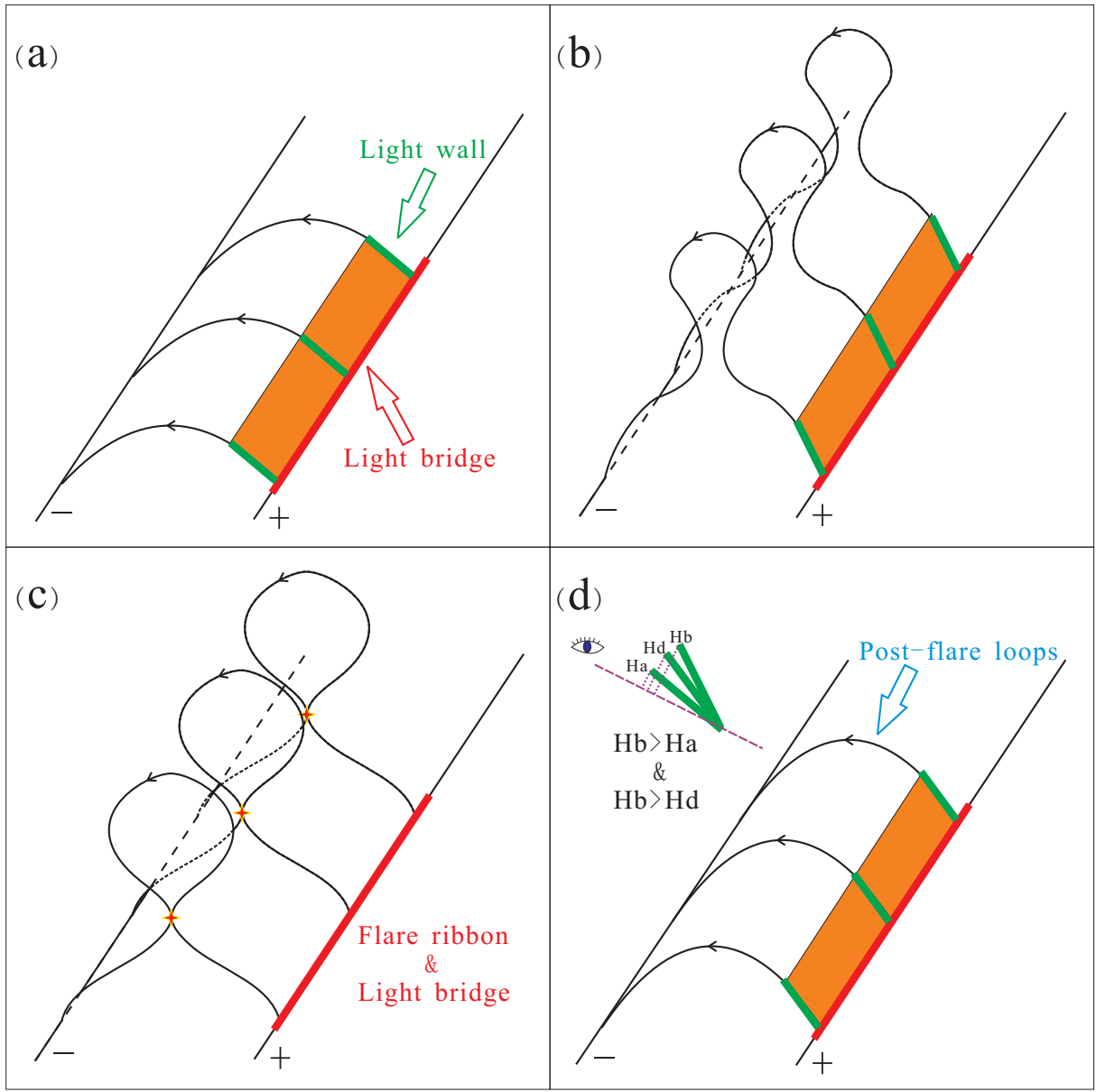}
\caption{
Schematic diagrams illustrating this study event.
The black straight lines mark the locations of the flare ribbons, while the plus (+) and minus
(-) symbols denote the polarities of the underneath magnetic fields. The black curves delineate
the magnetic field lines, which connect the two flare ribbons. The orange regions outline the
light wall above the light bridge. The red star symbols in panel (c) mark the reconnection sites.
In panel (d), we compare the mean projective maximum heights of the light wall at different stages.
``Ha" represents the mean projective maximum height of the light wall before the flare's start. It
increases to ``Hb" when the magnetic field lines are pushed upwards. After the magnetic reconnection
takes place, the mean projective maximum height of the light wall decreases from ``Hb" to ``Hd".
}
\label{fig4}
\end{figure}

\end{document}